# DMFSGD: A Decentralized Matrix Factorization Algorithm for Network Distance Prediction

Yongjun Liao, Wei Du, Pierre Geurts and Guy Leduc

***Abstract*—The knowledge of end-to-end network distances is essential to many Internet applications. As active probing of all pairwise distances is infeasible in large-scale networks, a natural idea is to measure a few pairs and to predict the other ones without actually measuring them. This paper formulates the distance prediction problem as matrix completion where unknown entries of an incomplete matrix of pairwise distances are to be predicted. The problem is solvable because strong correlations among network distances exist and cause the constructed distance matrix to be low rank. The new formulation circumvents the well-known drawbacks of existing approaches based on Euclidean embedding.**

**A new algorithm, so-called Decentralized Matrix Factorization by Stochastic Gradient Descent (DMFSGD), is proposed to solve the network distance prediction problem. By letting network nodes exchange messages with each other, the algorithm is fully decentralized and only requires each node to collect and to process local measurements, with neither explicit matrix constructions nor special nodes such as landmarks and central servers. In addition, we compared comprehensively matrix factorization and Euclidean embedding to demonstrate the suitability of the former on network distance prediction. We further studied the incorporation of a robust loss function and of non-negativity constraints. Extensive experiments on various publicly-available datasets of network delays show not only the scalability and the accuracy of our approach but also its usability in real Internet applications.**

## I. Introduction

On large networks such as the Internet, many applications require the knowledge of end-to-end network distances in order to achieve Quality of Service (QoS) objectives. In the networking field, the distance between two network nodes is defined by the delay or latency between them, in the form of either one-way delay or more often round-trip time (RTT). Examples include peer-to-peer file sharing, overlay networks, and content distribution systems where peers preferably access nodes or servers that are likely to respond fast [1]–[7].

Clearly, it is infeasible to probe actively end-to-end distances among all pairs of nodes in large networks as the demand in terms of measurements grows quadratically with the scale of the network. A natural idea is to probe a small set of pairs and then predict the distances between other pairs where there are no direct measurements. This understanding has motivated numerous research on Network Coordinate System (NCS) [8]–[12]. For instance, approaches based on Euclidean embedding have been widely studied and achieved good performance in interesting scenarios [9], [13]. Realizing that the assumption of Euclidean distance properties (symmetry and triangle inequality) are often violated in practice, as observed in various studies [9], [14]–[18], matrix factorization has recently drawn increasing attention of the networking community [10], [11].

In this paper, we investigate matrix factorization for network distance prediction. In particular, we formulate the problem of network distance prediction as a matrix completion problem where a partially observed matrix is to be completed [19]–[21]. Here, the matrix contains distance measurements such as RTTs between network nodes with some of them known and the others unknown thus to be filled. Matrix completion is only possible if matrix entries are largely correlated, which certainly holds for network distances because Internet paths with nearby end nodes often overlap and share common bottleneck links. These redundancies among network paths cause the constructed distance matrix to be low rank, which will be empirically demonstrated for various RTT datasets.

Although numerous approaches to matrix completion have been proposed, many of which are based on low-rank matrix factorization [22]–[24], very few are directly applicable to network applications where decentralized processing of data is most of the time a necessity. In this paper, we propose a fully decentralized algorithm based on Stochastic Gradient Descent (SGD), which is founded on the stochastic optimization theory with nice convergence guarantees [25].

The so-called Decentralized Matrix Factorization by Stochastic Gradient Descent (DMFSGD) algorithm has two distinct features. First, it requires neither explicit constructions of matrices nor special nodes such as landmarks and central servers where measurements are collected and processed. Instead, by letting network nodes exchange messages with each other, matrix factorization is collaboratively and iteratively achieved at all nodes, with each node equally retrieving a number of distance measurements. Second, the algorithm is simple, with no infrastructure, and is computationally lightweight, containing only vector operations. These features make it suitable for dealing with practical problems, when deployed in real applications, such as measurement dynamics where network measurements vary largely over time and network churn where nodes join and leave a network frequently. Extensive experiments on various publicly-available RTT datasets show not only the scalability and the accuracy of our approach but also its usability in real Internet applications.

Yongjun Liao and Guy Leduc are with Research Unit in Networking (RUN), University of Liège, Belgium. Email: {yongjun.liao, guy.leduc}@ulg.ac.be.

Wei Du is with Intelligent and Interactive Systems (IIS), University of Innsbruck, Austria. Email: wei.du@uibk.ac.at.

Pierre Geurts is with Systems and Modeling, University of Liège, Belgium. Email: p.geurts@ulg.ac.be.

Our preliminary work on decentralized matrix factorization for network distance prediction was published in [11]. In the present paper, we make the following distinct contributions:

- Our previous approach in [11] was based on Alternating Least Squares (ALS), requiring each node to probe all local measurements simultaneously. In contrast, the new SGD-based approach allows each node to probe one measurement at a time, making the system more flexible. The new approach also addresses several issues arising when applied practically, including the difficult choice of the learning rate parameter and the consideration of passive distance acquisitions and dynamic measurements.
- We compare comprehensively matrix factorization and Euclidean embedding to reveal the suitability of matrix factorization. A unified view is provided which leads to a unified optimization framework to solve both of them.
- Two extensions of the current matrix factorization model are proposed, including the incorporation of a robust loss function and the introduction of constraints in the model to ensure the nonnegativity of the predicted distances. These extensions are found helpful in improving the accuracy of the prediction and require little modification to the algorithm with no additional computational cost.
- In addition, more extensive evaluations have been carried out to study not only the impact of the parameters but also the accuracy of our approach. In particular, we highlight the usability of our approach by simulations on real dynamic data collected from a peer-to-peer file sharing application, namely Azureus [2], [13].

The rest of the paper is organized as follows. Section II summarizes the related work on network distance prediction based on Euclidean embedding and matrix factorization. Section III introduces the formulation of network performance prediction as matrix completion and its resolution by low-rank matrix factorization. Section IV describes the decentralized matrix factorization algorithm based on Stochastic Gradient Descent. Section V discusses possible extensions to the current matrix factorization model. Section VI evaluates our approach on various publicly available datasets of RTT. Conclusions and future work are given in Section VII.

## II. Related Work

Among numerous works on network distance prediction, we only discuss and compare approaches based on Euclidean embedding and on matrix factorization due to their simplicity and generality. We refer the interested readers to [12] for a more detailed review of this field.

### A. Euclidean Embedding

A straightforward approach to network distance prediction is to embed network nodes into a metric space where each node is assigned a coordinate from which distances can be directly computed. Two representatives are Globe Network Positioning (GNP) [8] and Vivaldi [9].

GNP firstly proposed the idea of network embedding that relies on a small number of landmarks. Based on inter-landmark distance measurements, the landmarks are first embedded into a metric space such as Euclidean or spherical coordinate systems. Then, the ordinary nodes calculate their coordinates with respect to the landmarks. Vivaldi extended GNP in a decentralized manner by eliminating the landmarks. It simulates the network by a physical system of spring and minimizes its energy according to Hooke's law to find an optimal embedding.

In all metric spaces, distances undergo two important properties:

- Symmetry: $\mathrm{d}(A,B) = \mathrm{d}(B,A)$;
- Triangle Inequality: $\mathrm{d}(A,B) + \mathrm{d}(B,C) \geqslant \mathrm{d}(A,C)$.

However, network distances are not necessarily symmetric especially when represented by one-way delays [26], [27]. The bigger issue is the property of triangle inequality. Many studies have shown that the violations of triangle inequality (TIV) are widespread and persistent in current Internet [9], [14]–[18]. In the presence of TIVs, metric space embedding shrinks the long edges and stretches the short ones, degrading heavily the accuracy of the embedding. Figure 1 illustrates the idea of Euclidean embedding for network distance prediction and the impact of TIVs on the accuracy.

Without loss of generality, we focus on the simplest metric space, namely Euclidean coordinate systems, in the rest of the paper.

### B. Matrix Factorization

Alternatively, matrix factorization has also been used for network distance prediction (see Figure 2 for an illustration). The biggest advantage of matrix factorization is that it makes no assumption of Euclidean distance properties and thus can tolerate the widespread TIVs and the possible asymmetry in network distance spaces.

The first system based on matrix factorization was Internet Distance Estimation Service (IDES) [10] which has the same landmark-based architecture as GNP. IDES factorizes a small but full inter-landmark distance matrix, at a so-called information server, by using Singular Value Decomposition (SVD). Similarly, Phoenix treated the early-entered nodes as landmarks and allowed an ordinary node to select any existing nodes in the system which already have coordinates assigned [28]. Landmark-based systems suffers from several drawbacks including single-point failures, landmark overloads, and potential security problems. The selection of landmarks can also affect the accuracy of the prediction. Moreover, in Section III-D, we will show that landmark-based approaches are actually special cases of a general decentralized matrix factorization model and thus can also be solved by our approach.

## III. Network Distance Prediction by Matrix Factorization

This section formulates the problem of network distance prediction as matrix completion and describes its resolution by matrix factorization. We also provide a unified view of different approaches to network distance prediction, the insights of which lead to a unified optimization framework.

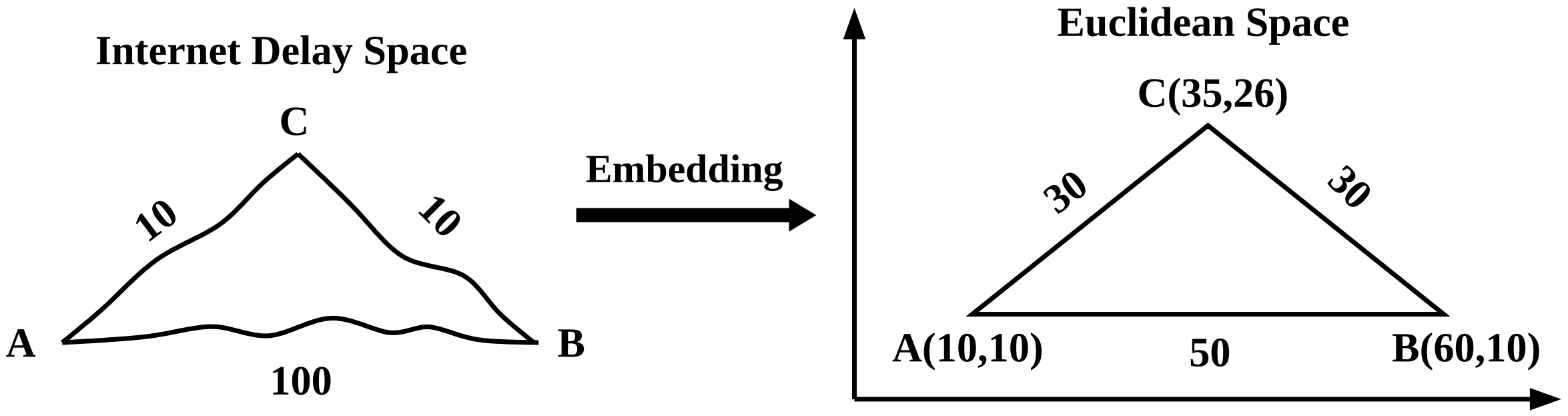

Fig. 1. Network distance prediction by Euclidean Embedding.

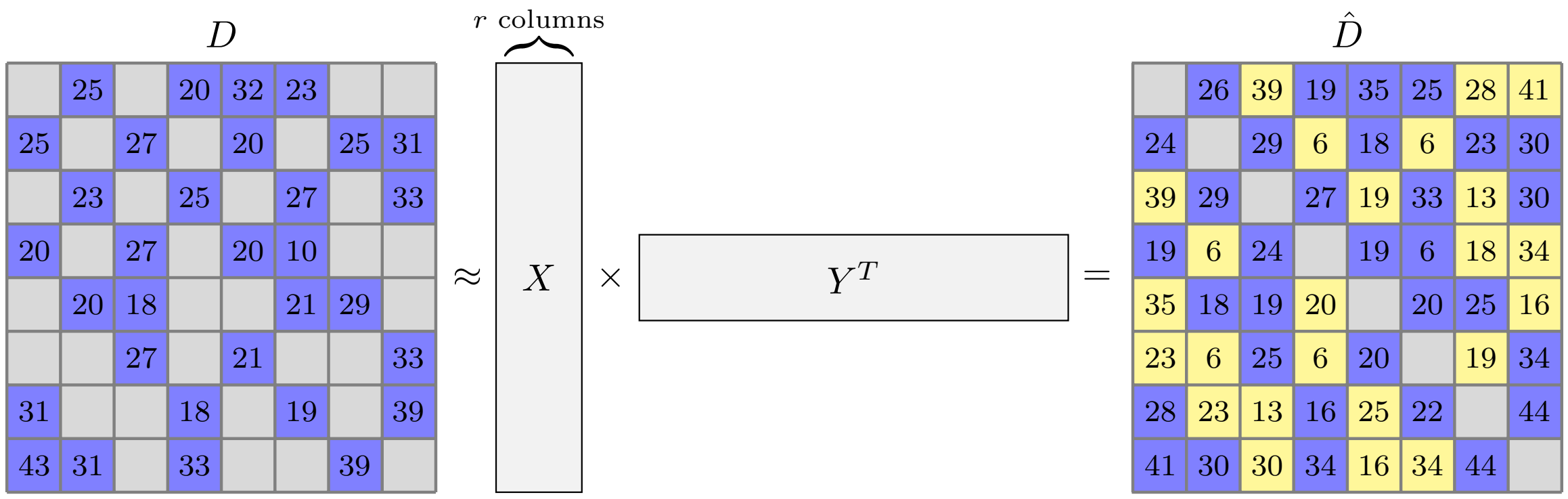

Fig. 2. Network distance prediction by matrix factorization. Note that the diagonal entries of $D$ and $\hat{D}$ are empty.

### *A. Problem Formulation*

Assuming $n$ nodes in the network, a $n \times n$ distance matrix is constructed with some distances between nodes measured and the others unmeasured. Denote $D$ the measured distance matrix with $d_{ij}$ the measured distance from node $i$ to node $j$ and $\hat{D}$ the predicted distance matrix with $\hat{d}_{ij}$ the predicted distance computed from some function.

Given the above notations, network distance prediction can be viewed as a matrix completion problem that estimates the missing entries in $D$ from a small number of known entries [20]. Its resolution generally amounts to minimizing a loss function of the following form

$$L(D, \hat{D}, W) = \sum_{i,j=1}^{n} w_{ij} l(d_{ij}, \hat{d}_{ij}), \tag{1}$$

where $W$ is a weight matrix with $w_{ij}$ taking values between 0 and 1. In a simple case, $w_{ij} = 1$ if $d_{ij}$ is measured and 0 otherwise. Note that if the distance measurements are RTTs, then $d_{ji} = d_{ij}$ as RTTs are approximately symmetric. Consequently, $w_{ji} = w_{ij}$ as $d_{ji}$ and $d_{ij}$ are either both known or both unknown.

$l$ is a loss function that penalizes the difference between an estimate and its desired or true value. The most commonly-used loss function is the $L_2$ or square loss function,

$$l(d, \hat{d}) = (d - \hat{d})^2. \tag{2}$$

We will discuss other loss functions in Section V.

### *B. Low-Rank Approximation and Matrix Factorization*

Additional constraints are needed to solve the matrix completion problem in eq. 1. A common approach is to constrain the rank of the approximate matrix $\hat{D}$ so that

$$Rank(\hat{D}) = r, \tag{3}$$

where $r \ll n$ for $D$ of size $n \times n$

The assumption in this low-rank approximation is that the entries of $D$ are largely correlated, which causes $D$ to have a low effective rank. To show that it holds for our problem, Figure 3 plots the singular values of two RTT matrices. It can be seen that the singular values of both matrices decrease fast as the 10th singular values are 5.7% and 2.9% of the largest ones respectively, indicating strong correlations in them. The low-rank nature of many other RTT datasets have been previously reported in [29].

Directly finding $\hat{D}$ by minimizing eq. 1 subject to eq. 3 is considerably difficult due to the rank constraint. However, as $\hat{D}$ is of low rank, we can factorize it into the product of two smaller matrices, i.e.,

$$\hat{D} = XY^T, \tag{4}$$

where $X$ and $Y$ are of size $n \times r$. Therefore, we can get rid of the rank constraint by replacing $\hat{D}$ by $XY^T$ in eq. 1, and then look for $X$ and $Y$ instead by minimizing

$$L(D, X, Y, W) = \sum_{i,j=1}^{n} w_{ij} l(d_{ij}, x_i y_j^T), \tag{5}$$

where $x_i$ is the $i$th row of $X$, $y_i$ is the $i$th row of $Y$, and $x_i y_j^T = \hat{d}_{ij}$ is the estimate of $d_{ij}$. Note that the factorization

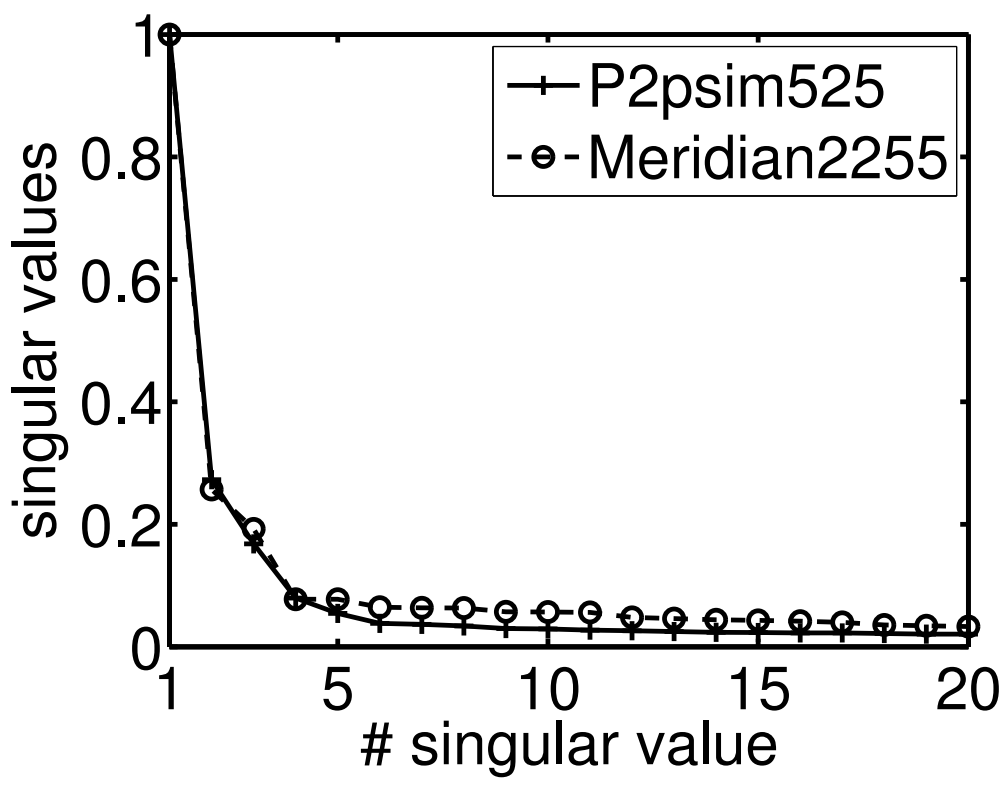

Fig. 3. The singular values of a RTT matrix of $2255 \times 2255$, extracted from the Meridian dataset [30] and called "Meridian2255", and of a RTT matrix of $525 \times 525$, extracted from the P2psim dataset [30] and called "P2psim525". The singular values are normalized so that the largest singular values of both matrices are equal to 1.

in eq. 4 has no unique solution as

$$\hat{D} = XY^T = XGG^{-1}Y^T, \tag{6}$$

where $G$ is any arbitrary $r \times r$ invertible matrix. Thus, replacing $X$ by $XG$ and $Y^T$ by $G^{-1}Y^T$ results in the same $\hat{D}$.

Generally, the class of techniques to solve the low-rank approximation is matrix factorization. When $D$ is complete, analytic solutions can be found by using singular value decomposition (SVD) [31]. With missing entries, the factorization is usually done by iterative optimization methods such as Gradient Descent or Newton algorithms [32]. Note that additional constraints can be imposed in eq. 5. For instance, the entries of $X$ and $Y$ can be required to be nonnegative in order to recover a nonnegative matrix, leading to the nonnegative matrix factorization (NMF) [33].

### C. Incorporation of the Regularization

Matrix completion by matrix factorization suffers from a well-known problem called overfitting in the field of machine learning [34]. In words, directly optimizing eq. 5 often leads to a "perfect" model with no or small errors on the training data while having large errors on the unseen data which are not used in learning. The problem is more severe when $D$ is sparse or when $r$ is large.

A common way to avoid overfitting is through regularization that penalizes the norms of the solutions, resulting in the following regularized loss function,

$$L(D, X, Y, W, \lambda) = \sum_{i,j=1}^{n} w_{ij} l(d_{ij}, x_i y_j^T) + \lambda \sum_{i=1}^{n} x_i x_i^T + \lambda \sum_{i=1}^{n} y_i y_i^T, \tag{7}$$

where $\lambda$ is the regularization coefficient that controls the extent of regularization.

Besides avoiding overfitting, the regularization also helps overcome the drift of the solutions due to the non-uniqueness of the factorization (see eq. 6), which often leads to the overflows of the solutions. Among the infinite number of pairs of $X$ and $Y$ which produce the same $\hat{D}$, the incorporation of the regularization will force to choose the pair with the smallest norm.

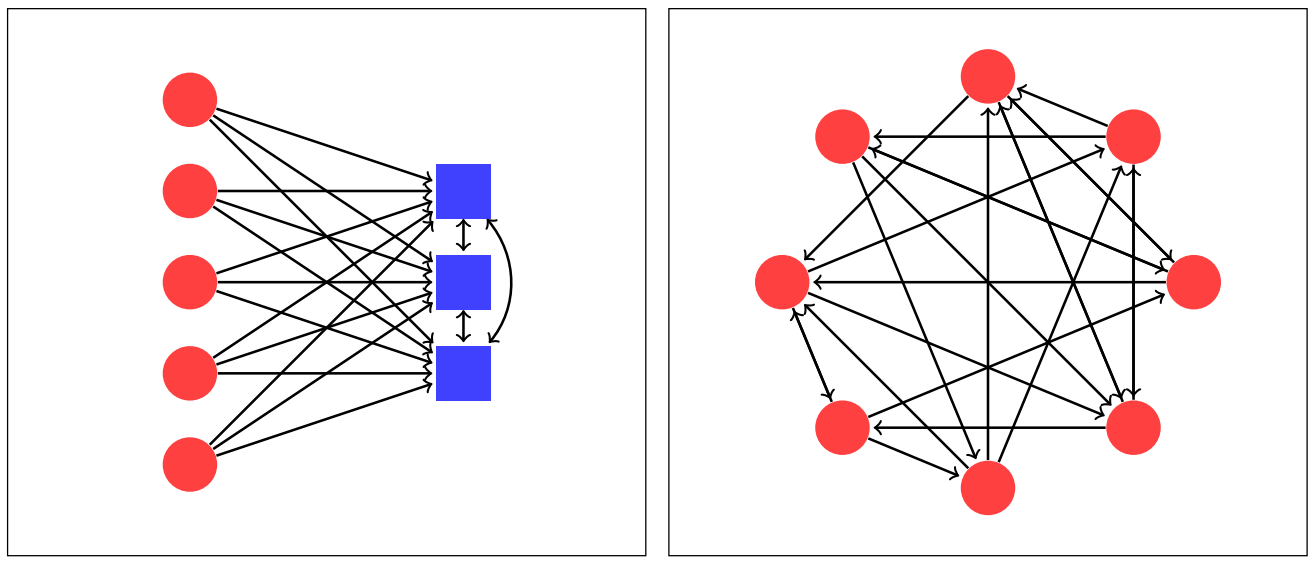

Fig. 4. Architectures of landmark-based, the left plot, and decentralized, the right plot, systems for network distance prediction. The squares are landmarks and the circles are ordinary nodes. The directed path from node $i$ to node $j$ means that node $i$ probes node $j$ and therefore $w_{ij} = 1$.

### D. A Unified View of Approaches to Network Distance Prediction

Although popular approaches to network distance prediction vary by adopting various models including Euclidean embedding and matrix factorization and by adopting different architectures of either landmark-based or landmark-less and thus decentralized, these seemingly different approaches all optimize the same function in eq. 1 but differ only in the setting of $w_{ij}$ and in the associated distance functions to calculate $\hat{d}_{ij}$.

- Setting of $w_{ij}$: For landmark-based methods, as all paths between landmarks are measured and ordinary nodes probe only the landmarks,
$$w_{ij} = \begin{cases} 1 & \text{if node } j \text{ is a landmark} \\ 0 & \text{otherwise} \end{cases}.$$
For decentralized methods, as each node equally probes a number of nodes,
$$w_{ij} = \begin{cases} 1 & \text{if node } i \text{ probes node } j \\ 0 & \text{otherwise} \end{cases}.$$
Figure 4 illustrates the architectures of landmark-based and decentralized systems.
- Distance functions to calculate $\hat{d}_{ij}$: For matrix factorization, as described above,
$$\hat{d}_{ij} = x_i y_j^T, \tag{8}$$
For Euclidean embedding, the Euclidean distance is defined as
$$\hat{d}_{ij} = \sqrt{(x_i - x_j)^T (x_i - x_j)}, \tag{9}$$
where $x_i$ and $x_j$ are the Euclidean coordinates of node $i$ and node $j$.

The above insights suggest a unified framework to treat and to solve equally network distance prediction under different models and different architectures. For instance, the decentralized matrix factorization algorithms proposed in the following sections can be used to solve both Euclidean embedding and landmark-based systems with little modification.

# IV. Decentralized Matrix Factorization for Network Distance Prediction

The goal is to find $X$ and $Y$ such that $XY^T$ best approximates $D$ by minimizing Eq. 7. Commonly, it requires to collect and to process a number of distance measurements at a central node, which is a major obstacle to network applications. Below, we introduce algorithms to minimize eq. 7 in a fully decentralized manner.

## A. Problem Formulation

A decentralized resolution of eq. 7 forbids the explicit constructions of large matrices. Thus, each row of $X$ and of $Y$, denoted by $x_i$ and $y_i$ and called the $x$ and $y$ coordinates of node $i$ in the sequel, are stored distributively at each node. To calculate $x_i$ and $y_i$ locally, we let node $i$ probe and exchange messages with a number of nodes in the network, called *neighbors* in the sequel, and optimize the following losses

$$l_i = \sum_{j=1}^{n} w_j^i l(d_{ij}, x_i y_j^T) + \lambda x_i x_i^T, \tag{10}$$

$$l^i = \sum_{j=1}^{n} w_j^i l(d_{ji}, x_j y_i^T) + \lambda y_i y_i^T, \tag{11}$$

where $w_j^i$ represents the neighboring relationship of node $j$ to node $i$, i.e., $w_j^i = 1$ if node $j$ is a neighbor of node $i$ or equivalently if node $i$ probes node $j$, and 0 otherwise. Note that as $w_j^i$ is not necessarily equal to $w_i^j$, measurements $d_{ij}$ and $d_{ji}$ may only be known by one node of either node $i$ or node $j$ but not by the other.

Essentially, $l_i$ is the regularized loss of the edges from node $i$ to other nodes and $l^i$ is that of the edges from other nodes to node $i$. Thus, we address the large-scale optimization problem in eq. 7 by decomposing it into a number of subproblems in eqs. 10 and 11 which can be solved locally at each node by using only local measurements.

In seeing that eqs. 10 and 11 are standard least-squares problems where analytical solutions exist, our previous work in [11] solved the matrix factorization problem by Alternating Least Squares (ALS), which alternatively and iteratively solves the small least-squares problems in eqs. 10 and 11. While the ALS-based algorithm performed well in simulations on datasets containing static measurements, it requires each node to probe measurements with a number of nodes simultaneously, which is impractical when deployed in real applications. Below, we propose a different algorithm based on Stochastic Gradient Descent (SGD) that processes, at each node, measurements one by one and one at a time.

## B. Stochastic Gradient Descent (SGD)

SGD is a variation of traditional Batch Gradient Descent which is often used for online machine learning [25]. Instead of collecting all training samples beforehand and computing the gradients over them, each iteration of SGD chooses one training sample at random and updates the parameters being estimated along the negative gradients computed over that chosen sample. SGD is particularly appropriate for network applications, as measurements can be acquired on demand and processed locally at each node. It also has simple update rules that involve only vector operations and is able to deal with large-scale dynamic measurements.

*1) Stochastic Updates:* When using SGD, each node probes one neighbor at a time, measures its distance with respect to that node and retrieves that node's coordinates. Let node $j$ be the chosen neighbor by node $i$ at the current time. Then, the regularized losses that node $i$ seeks to reduce with respect to node $j$ are

$$l_{ij} = l(d_{ij}, x_i y_j^T) + \lambda x_i x_i^T, \tag{12}$$

$$l_{ji} = l(d_{ji}, x_j y_i^T) + \lambda y_i y_i^T. \tag{13}$$

The gradients of $l_{ij}$ and $l_{ji}$ are

$$\frac{\partial l_{ij}}{\partial x_i} = \frac{\partial l(d_{ij}, x_i y_j^T)}{\partial x_i} + \lambda x_i, \tag{14}$$

$$\frac{\partial l_{ji}}{\partial y_i} = \frac{\partial l(d_{ji}, x_j y_i^T)}{\partial y_i} + \lambda y_i. \tag{15}$$

In particular, the gradients of the $L_2$ loss function are

$$\frac{\partial l}{\partial x_i} = -(d_{ij} - x_i y_j^T) y_j, \tag{16}$$

$$\frac{\partial l}{\partial y_i} = -(d_{ji} - x_j y_i^T) x_j. \tag{17}$$

Note that we dropped the factor 2 from the derivatives of the regularization terms and of the $L_2$ loss function for mathematical convenience.

Then, node $i$ updates its coordinates along the negative gradient directions, given by

$$x_i = (1 - \eta\lambda) x_i + \eta(d_{ij} - x_i y_j^T) y_j, \tag{18}$$

$$y_i = (1 - \eta\lambda) y_i + \eta(d_{ji} - x_j y_i^T) x_j, \tag{19}$$

where $\eta$, called *learning rate* or *step size*, controls the speed of the updates.

*2) Minibatch and Line Search:* The SGD algorithm is sensitive to the learning rate $\eta$, where a too large $\eta$ results in large steps of updates and may overflow the solution, whereas a too small $\eta$ makes the convergence slow. This sensitivity can be relieved by using more training samples at the same time, leading to minibatch SGD with the following update rules

$$x_i = (1 - \eta\lambda) x_i + \eta \sum_{j=1}^{n} w_j^i (d_{ij} - x_i y_j^T) y_j, \tag{20}$$

$$y_i = (1 - \eta\lambda) y_i + \eta \sum_{j=1}^{n} w_j^i (d_{ji} - x_j y_i^T) x_j. \tag{21}$$

To completely get rid of $\eta$, a line search strategy can be incorporated to determine $\eta$ adaptively [35]. In particular, in each update, $\eta$ starts with a large initial value and is gradually decreased until the losses in eqs. 10 or 11 are reduced after the update. The line search algorithm for updating $x_i$ is given in Algorithm 1. The same algorithm can be used for updating $y_i$ by replacing eq. 10 by eq. 11 and eq. 20 by eq. 21. Note that $\delta$ in Line 6 is a small positive constant that helps overcome poor local optimums. We will demonstrate the effectiveness of adapting $\eta$ by line search using Algorithm 1 in Section VI.

**Algorithm 1** Line Search (for updating $x_i$)

compute $l_i^0$ by eq. 10;
initialize $\eta$ with a large value;
**for** $i = 1$ to $maxNumberLineSearch$ **do**
  compute $x_i$ by eq. 20;
  compute $l_i$ by eq. 10;
  **if** $l_i < l_i^0 + \delta$ **then**
    **return**
  **end if**
  $\eta \longleftarrow \eta/2$;
**end for**

### C. Neighbor Decay and Neighbor Selection

As mentioned earlier, it is preferable to have a system that probes and processes measurements one by one. Thus, we let each node maintain the information (distance measurements and coordinates) of its neighbors, i.e., the nodes with which it communicates. In minibatch SGD, each node probes one neighbor at a time but updates its coordinates with respect to all neighbors in the neighbor set using their recorded historical information.

A neighbor decay strategy is incorporated that scales the weight of each node in the neighbor set by its age so that older information receives less weight, i.e.,

$$w_j^i = \frac{a_{max} - a_j}{\sum_{j \in \text{NeighborSet}(i)} (a_{max} - a_j)}, \tag{22}$$

where $a_j$ is the age of the information of node $j$ and $a_{max}$ is the age of the oldest information in the neighbor set. Note that this neighbor decay strategy was firstly proposed by [13] to overcome the problem of skewed neighbor update in Vivaldi. In words, some nodes may be probed at far greater frequency than others due simply to their longer life cycles and a direct consequence is that the optimization will become skewed toward these nodes.

Conventionally, the neighbors of a node are selected randomly and the distances between a node and its neighbors are probed by active measurements [9]. However, in practice, it is more attractive to perform the updates of the coordinates passively without generating any extra traffic. In some applications such as Azureus, passivity is enforced, as we have no control over the selection of neighbors with which a node communicates and when it communicates with them [13].

Therefore, we differentiate the situations where distances are probed by active and passive measurements. For the former, the conventional random neighbor selection procedure is adopted, i.e., each node randomly selects $k$ nodes as its neighbors and actively probes one of them from time to time. For the latter, no neighbor selection is performed explicitly and each node maintains a relatively small set of active neighbors with which it recently communicated and updates its coordinates whenever a new measurement is made available. Note that this difference has no impact on the update rules in eqs 18 and 19 or in eqs 20 and 21.

**Algorithm 2** DMFSGD($i$, $j$)

node $i$ retrieves $d_{ij}, d_{ji}, x_j, y_j$ actively or passively;
node $i$ updates the weights of its neighbors by eq. 22;
update $x_i$ by eq. 20 with $\eta$ set by line search;
update $y_i$ by eq. 21 with $\eta$ set by line search;

### D. Algorithm

We denote the SGD-based decentralized matrix factorization algorithm as DMFSGD, given in Algorithm 2. Like Vivaldi [9], our DMFSGD algorithm has no infrastructure and employs the same process at all nodes. It is simple, with update rules containing only vector operations.

In the implementation, the coordinates of each node are initialized with random numbers uniformly distributed between 0 and 1. Empirically, the algorithm is insensitive to the random initializations of the coordinates. We would like to point out that the algorithm is one of those randomized gossip algorithms where each node exchanges messages with a number of other nodes randomly [36].

As mentioned earlier, the algorithm is generic and can also deal with landmark-based architectures, by letting each node only select landmarks as its neighbors, and with Euclidean embedding, by adopting the Euclidean distance defined as eq. 9 when optimizing eq. 1, which leads to the update rule of Vivaldi [9], given by

$$x_i = x_i - \eta \frac{\partial l(d_{ij}, \hat{d}_{ij})}{\partial x_i} = x_i + \eta (d_{ij} - \hat{d}_{ij}) \frac{x_i - x_j}{\hat{d}_{ij}}.$$

Note that Vivaldi adopted the $L_2$ loss function in eq. 1 with no regularization incorporated, and the learning rate $\eta$, termed differently as *timestep*, was adapted by taking into account some confidence measure of each node to its coordinate. Thus, Vivaldi can be viewed as a SGD-based decentralized Euclidean embedding algorithm, instead of the simulation of a spring system in [9].

## V. Extended Matrix Factorization Models

This section discusses possible ways to extend the common matrix factorization model.

### A. Robust Matrix Factorization

The widely-used $L_2$ loss function is known to be sensitive to outliers which often occur in network measurements due to network anomaly such as sudden traffic bursts and attacks from malicious nodes. Other loss functions such as the $L_1$ loss function, the $\epsilon$-insensitive loss function and the Huber loss function are more robust and can tolerate outliers [37], [38]. For example, the $L_1$ loss function is defined as

$$l(d, \hat{d}) = |d - \hat{d}|. \tag{23}$$

Thus, we can potentially enhance the robustness of matrix factorization by replacing the $L_2$ loss function by e.g. the $L_1$ loss function, and the same SGD procedure can be applied to solve the robust matrix factorization problem. Note that the $L_1$ loss function is non-differentiable and the gradients have to be approximated by the subgradients [1], given by

$$\frac{\partial l}{\partial x_i} = -\text{sign}(d_{ij} - x_i y_j^T) y_j, \tag{24}$$

$$\frac{\partial l}{\partial y_i} = -\text{sign}(d_{ji} - x_j y_i^T) x_j. \tag{25}$$

Replacing the gradient functions of eqs. 14 and 15 by eqs. 24 and 25, the update rules of minibatch SGD become

$$x_i = (1 - \eta\lambda) x_i + \eta \sum_{j=1}^{n} \text{sign}(d_{ij} - x_i y_j^T) w_j^i y_j, \tag{26}$$

$$y_i = (1 - \eta\lambda) y_i + \eta \sum_{j=1}^{n} \text{sign}(d_{ji} - x_j y_i^T) w_j^i x_j, \tag{27}$$

Comparing eqs. 26 and 27 with eqs. 20 and 21, the only difference is that for the $L_2$ loss function, the magnitudes of the updates are proportional to the fitting errors $(d - xy^T)$, whereas for the $L_1$ loss function, only the signs of the fitting errors are taken into consideration and decide the directions of the updates.

### B. NonNegativity Constraint

Conventional matrix factorization techniques do not preserve the nonnegativity of the distances. Empirically, only a very small portion of the predicted distances were found negative by our DMFSGD algorithm, and a direct solution is to turn $\hat{d}_{ij}$ into a small positive value if $\hat{d}_{ij} = x_i y_j^T < 0$.

A systematic solution is to incorporate the nonnegativity constraint in matrix factorization, leading to the nonnegative matrix factorization (NMF) that optimizes

$$\sum_{i,j=1}^{n} w_{ij} l(d_{ij}, x_i y_j^T) + \lambda \sum_{i=1}^{n} x_i x_i^T + \lambda \sum_{i=1}^{n} y_i y_i^T, \tag{28}$$
$$\text{subject to } x_i \geqslant 0,\ y_i \geqslant 0,\ i = 1, \ldots, n.$$

The optimization of NMF is not fundamentally different from that of the unconstrained matrix factorization, adding only one projection step that turns the negative entries in $x_i$ and $y_i$ into zero after each SGD update which causes no noticeable impact on the speed of the algorithm. The technique is also known as projected gradient descent [39].

Note that the nonnegativity constraint has been previously studied in [10], [28], both of which adopted a more heavy-weight nonnegative least-squares solver.

### C. Symmetric Distance Matrix Factorization

Also note that network distances are symmetric if represented by RTT and that this symmetry is not preserved either. A direct solution is to turn the predicted distances symmetric by defining a symmetric distance function as

$$\hat{d}_{ij}^s = \frac{\hat{d}_{ij} + \hat{d}_{ji}}{2} = \frac{x_i y_j^T + x_j y_i^T}{2}. \tag{29}$$

[1]Analogously, the subgradient-based technique that optimizes non-differentiable functions is called subgradient descent [35]. Following the convention in [25], we use the term SGD to refer to both Stochastic Gradient and SubGradient Descent.

**Algorithm 3** Extended_DMFSGD($i$, $j$)

node $i$ retrieves $d_{ij}, d_{ji}, x_j, y_j$ actively or passively;
node $i$ updates the weights of its neighbors by eq. 22;
**if** use $L_2$ loss function **then**
  update $x_i$ by eq. 20 with $\eta$ set by line search;
  update $y_i$ by eq. 21 with $\eta$ set by line search;
**else** // use $L_1$ loss function
  update $x_i$ by eq. 26 with $\eta$ set by line search;
  update $y_i$ by eq. 27 with $\eta$ set by line search;
**end if**
**if** force nonnegativity **then**
  turn the negative entries in $x_i$ and $y_i$ into 0;
**end if**

As distances are defined as in eq. 29, a systematic solution is to factorize $D$ by optimizing

$$\sum_{i=1}^{n} \sum_{j=1}^{n} w_{ij} l(d_{ij}, \hat{d}_{ij}^s) + \lambda \sum_{i=1}^{n} x_i x_i^T + \lambda \sum_{i=1}^{n} y_i y_i^T. \tag{30}$$

Similar SGD update rules can be derived.

### D. Height Model

The height model in Vivaldi [9] can also be incorporated. This model augments the $x$ and $y$ coordinates of a node with a height. Similarly, the $x$ and $y$ coordinates model the high-speed Internet core, while the height models the time packets take to travel the access link from the node to the Internet core. The cause of the access link distance includes queuing delay and low bandwidth [9]. The height augmented symmetric distance is defined as

$$\hat{d}_{ij}^{hs} = \frac{x_i y_j^T + x_j y_i^T}{2} + h_i + h_j. \tag{31}$$

Correspondingly, the loss function to be optimized becomes

$$\sum_{i=1}^{n} \sum_{j=1}^{n} w_j^i l(d_{ij}, \hat{d}_{ij}^{hs}) + \lambda \sum_{i=1}^{n} x_i x_i^T + \lambda \sum_{i=1}^{n} y_i y_i^T. \tag{32}$$

Similar SGD update rules can be derived.

### E. Extended DMFSGD Algorithm

Empirically, we found no or little improvements by incorporating the symmetric or height-augmented symmetric distance function in eq. 29 or 31, thus include neither of them in our system. However, the incorporation of the nonnegativity constraint and the robust loss function not only improved the accuracy but also made the results more stable and less sensitive to parameter settings, which will be demonstrated in Section VI. The extended DMFSGD algorithm is given in Algorithm 3. Note that as the basic version in Algorithm 2 is a special case of the extended version in Algorithm 3, we will refer to Algorithm 3 simply as DMFSGD in the sequel.

## VI. Experiments and Evaluations

In this section, we evaluate our DMFSGD algorithm[2] and compare it with state-of-the-art approaches.

[2]The source code of the algorithm will be publicly available soon.

### A. Evaluation Methodology

The evaluations were performed under the following criteria and on the following datasets.

*1) Evaluation Criteria:*

- **Cumulative Distribution of Relative Estimation Error** Relative Estimation Error (REE) is defined as
$$REE = \frac{|\hat{d}_{ij} - d_{ij}|}{d_{ij}}.$$
- **Stress** Stress measures the overall fitness and is used to illustrate the convergence of the algorithm, defined as
$$stress = \sqrt{\frac{\sum_{i,j=1}^{n}(d_{ij} - \hat{d}_{ij})^2}{\sum_{i,j=1}^{n} {d_{ij}}^2}}.$$
- **Median Absolute Error** Median Absolute Error (MAE) is defined as
$$MAE = median_{ij}(|d_{ij} - \hat{d}_{ij}|).$$

*2) Datasets:*

- **Harvard226** contains dynamic and passive measurements of application-level RTTs, with timestamps, between 226 Azureus clients collected in 4 hours [13].
- **P2PSim1740** was obtained from the P2PSim project that contains static RTT measurements between 1740 Internet DNS servers [40], [41].
- **Meridian2500** was obtained from the Cornell Meridian project that contains static RTT measurements between 2500 nodes [30].
- **P2PSim525** is a complete submatrix between 525 nodes derived from P2psim1740.
- **Meridian2255** is a complete submatrix between 2255 nodes derived from Meridian2500.
- **Synthetic1000** contains the pairwise distances between 1000 nodes that are randomly generated in a 10-dimensional Euclidean space.

The first five datasets were obtained from real-world networks and contain a large percentage of TIV edges, whereas the last one was synthesized and is TIV free. Here, an edge $\overline{AB}$ is claimed to be a TIV if there exists a triangle $\triangle ABC$ where $\overline{AB} > \overline{BC} + \overline{AC}$. The last three datasets were only used in section VI-B for the purpose of comparing the models of Euclidean embedding and matrix factorization.

Table I summarizes these datasets. Note that we can neither tell the symmetry nor calculate the TIV percentage of the Harvard226 dataset, as the measurements between network nodes vary over time largely, sometimes in several orders of magnitudes. The Harvard226 dataset is rather dense with about 3.9% pairwise paths unmeasured in 4 hours. The other paths are measured in uneven frequencies with one measured the maximal 662 times and one the minimal 2 times. About 94.0% of the paths are measured between 40 and 60 times.

TABLE I
PROPERTIES OF THE DATASETS

| Dataset | Nodes | Symmetry | TIV percentage | Dynamic |
|---|---|---|---|---|
| Harvard226 | 226 | / | / | Yes |
| P2PSim1740 | 1740 | Yes | 85.53% | No |
| Meridian2500 | 2500 | Yes | 96.55% | No |
| P2PSim525 | 525 | Yes | 76.17% | No |
| Meridian2255 | 2255 | Yes | 96.25% | No |
| Synthetic1000 | 1000 | Yes | No | No |

*3) Implementations for Different Datasets:* As mentioned earlier, the DMFSGD algorithm adopts the conventional random neighbor selection procedure in the scenarios where measurements are probed actively and maintains dynamically an active neighbor set for each node in the scenarios where measurements are obtained passively. Thus, for the Harvard226 dataset, we let each node maintain an active neighbor set containing the nodes it has contacted within the past 30 minutes and the timestamped measurements are processed in time order. For the other datasets, the random neighbor selection is used and the measurements are processed in random order with no neighbor decay (Line 2 in Algorithm 3) as they are static.

To handle the dynamics of the measurements in Harvard226, the distance filter in [13] is adopted that smooths the streams of measurements within a moving time window, 30 minutes in this paper, by a median filter. In the evaluation, we built a static distance matrix by extracting the median values of the streams of measurements between each pair of nodes and used it as the ground truth.

### B. Euclidean Embedding vs. Matrix Factorization

Euclidean embedding and matrix factorization both solve the same problem in eq. 1 but subject to different constraints. Euclidean embedding requires $\hat{D}$ to be symmetric and to satisfy the triangle inequality, whereas matrix factorization only requires $\hat{D}$ to be low rank. Below, we compare empirically Euclidean embedding and matrix factorization to show whether this difference in constraints makes matrix factorization more suitable for network distance prediction.

*1) Algorithms:* To make the model comparison fair, we chose the state-of-the-art algorithms to solve the Euclidean embedding and matrix factorization problems so that both are solved to their limits. For Euclidean embedding, Multi-Dimensional Scaling (MDS) is the most popular technique that searches the optimal embedding using an iterative algorithm. We adopted the MDS implementation, mdscale, in the statistical toolbox of matlab [42].

For matrix factorization, SVD provides the analytic solution which is globally optimal [31]. Generally, SVD factorizes a given matrix $D$ into three matrices of the form

$$D = USV^T,$$

where $U$ and $V$ are unitary matrices, and $S$ is a diagonal matrix with nonnegative real numbers on the diagonal. The positive diagonal entries are called the singular values and their number is equal to the rank of $D$.

To obtain a low-rank factorization, we keep only the $r$ large singular values in $S$ and replace the other small ones by zero. Let $S_r$ be the new $S$, $X = US_r^{\frac{1}{2}}$ and $Y^T = S_r^{\frac{1}{2}}V^T$, where

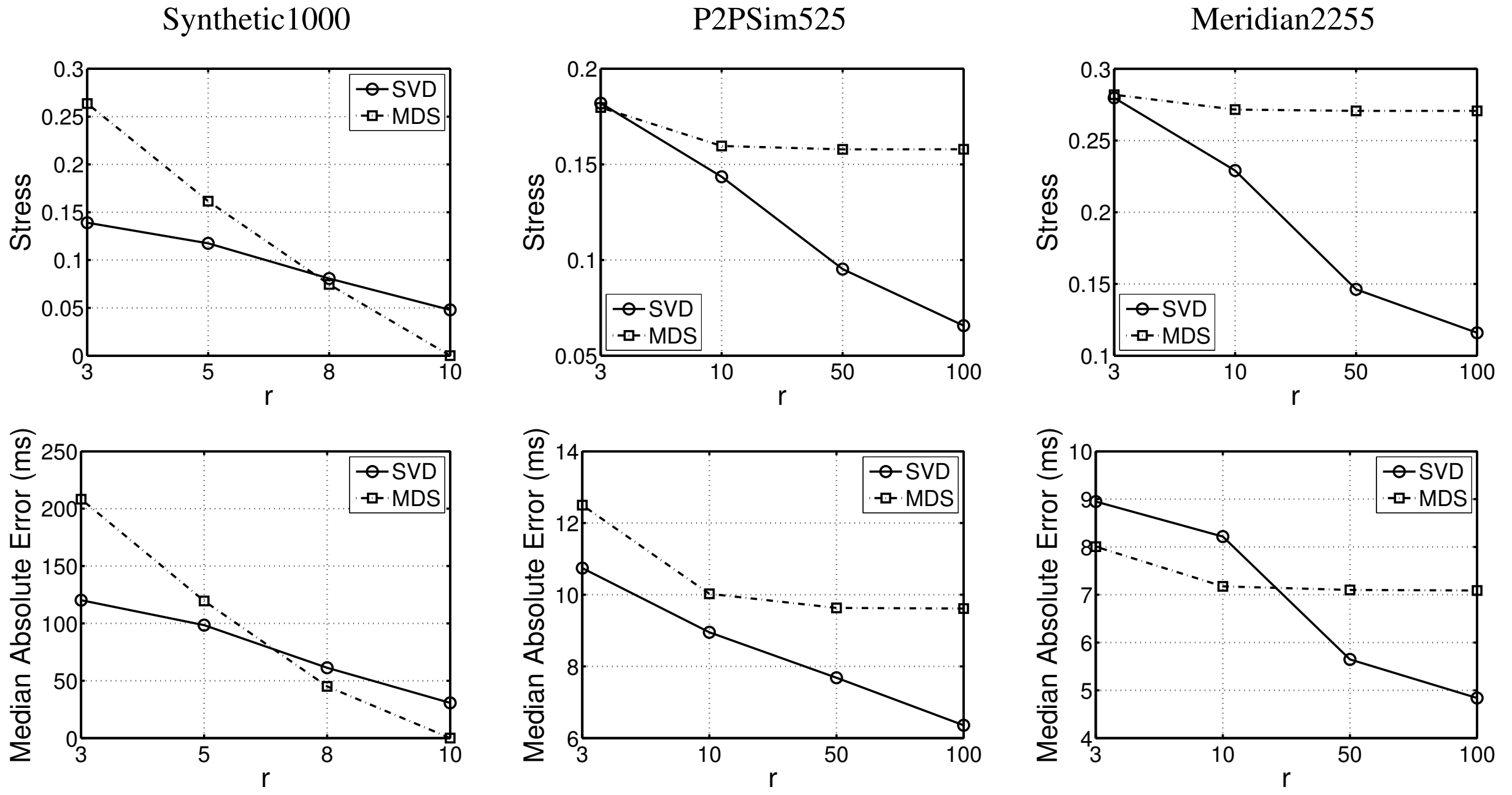

Fig. 5. Comparison of MDS-based Euclidean embedding and SVD-based matrix factorization on synthetic1000, P2psim525 and Meridian2255. The stresses and the median absolute errors by both methods in different dimensions/ranks are shown on the first and second rows respectively. Note that a perfect embedding with no errors was generated for Synthetic1000 in the 10 dimensional Euclidean space by MDS.

TABLE II
MATRIX FACTORIZATION VS. EUCLIDEAN EMBEDDING

| | Matrix Factorization | Euclidean Embedding |
|---|---|---|
| node coordinate | $x_i = (x_{i1}, \cdots, x_{ir})$ <br> $y_i = (y_{i1}, \cdots, y_{ir})$ | $x_i = (x_{i1}, \cdots, x_{ir})$ |
| distance function | $\hat{d}_{ij} = x_i y_j^T$ | $\hat{d}_{ij} = \sqrt{(x_i - x_j)^T (x_i - x_j)}$ |
| constraints | low rank | Symmetry: $\hat{d}_{ij} = \hat{d}_{ji}$ <br> Triangle Inequality: $\hat{d}_{ij} < \hat{d}_{ik} + d_{kj}$ |

$S_r(i,i)^{\frac{1}{2}} = \sqrt{S_r(i,i)}$. Then, $\hat{D} = XY^T$ is the optimal low-rank approximation to $D$.

*2) Evaluations:* Since SVD cannot handle missing data, we only compare MDS and SVD on the three complete datasets including Synthetic1000, P2psim525 and Meridian2255. On each dataset, we ran MDS and SVD in different dimensions and ranks and computed the stresses and MAE, shown in figure 5. It can be seen that the accuracies by SVD monotonically improve on all three datasets as the rank increases, whereas consistent improvements by MDS are only found on Synthetic1000 which is TIV-free. On P2psim525 and Meridian2255 where severe TIVs exist, MDS achieves no or little visible improvement after 10 dimensions.

These evaluations demonstrate the influences of different constraints imposed on the two techniques. For Euclidean embedding, the symmetry constraint doesn't cause any problem as the RTTs in all datasets are symmetric. However, the constraint of triangle inequality is strong and can't be relieved by increasing dimensions. In contrast, matrix factorization makes no assumptions of triangle inequality, thus is not affected by the TIVs in the data. Note that the accuracy improvement by increasing the rank is guaranteed for SVD-based matrix factorization. However, this conclusion cannot be extended to the cases where missing data is present. We will show later that increasing the rank beyond some value in matrix factorization for a large amount of missing data will not further improve the accuracy.

This comparative study reveals the model advantages of Matrix Factorization over Euclidean embedding. Overall, Euclidean embedding has a geometric interpretation which is useful for visualization. However, due to the existence of TIVs and the possible asymmetry in network distance spaces, low-rank matrix factorization is more suitable for modeling the network distance spaces. Table II compares the main features of matrix factorization and Euclidean embedding.

### C. Impact of Parameters

This section discusses and demonstrates the impact of the parameters of our DMFSGD algorithm.

*1) $k$, $r$ and $\lambda$:* Our DMFSGD algorithm has two main parameters, the regularization coefficient $\lambda$ and the rank $r$. Active probing introduces one additional parameter, the number of neighbors $k$ that are selected for each node. Intuitively, $r$ is the number of unknown variables in each coordinate and $k$ is the amount of known data that is used to estimate each unknown variable in a coordinate. $\lambda$ controls the extent of the regularization which avoids both overfitting and drift of the coordinates.

Clearly, increasing $k$ is equivalent to adding more data and thus always helps improve the accuracy. However, a larger $k$ also means more probe traffic and consequently higher overheads. On the other hand, only a certain number of unknown variables can be accurately calculated from a certain amount of known data. Thus, increasing $r$ beyond some value for a fixed $k$ will only lead to severe overfitting and consequently, a large $\lambda$ is needed to address it.

In practice, $k$ should be fixed according to the requirements of the applications by trading off between accuracies and measurement overheads. Following the suggestion in Vivaldi [9], we set $k = 32$ for P2psim1740 and for Meridian2500 in the rest of the paper. Note that $k = 32$ makes the available measurements considerably sparse. For instance, $32/1740 = 1.84\%$ measurements are available for each node in P2PSim1740 and $32/2500 = 1.28\%$ for each node in Meridian2500. Recall that no $k$ is set for Harvard226.

*2) Experiments under Different Configurations:* We then experimented with different configurations of $r = \{3, 10, 100\}$ and $\lambda = \{0.01, 0.1, 1, 10\}$, with different loss functions and whether to incorporate the nonnegativity constraint, shown in Figure 6. $\eta$ is adapted by the line search, with the initial value of $10^{-3}$ for the $L_2$ loss function and of $10^{-2}$ for the $L_1$ loss function.

In particular, we made the following observations. First, the DMFSGD algorithm is generally more accurate when the robust $L_1$ loss function and the nonnegativity constraint are incorporated. The likely reasons are that the $L_1$ loss function is insensitive to large fitting errors some of which are introduced by measurement outliers and that the nonnegativity constraint reduces the searching space which makes it easier to find a stable solution. Thus, the robust $L_1$ loss function and the nonnegativity constraint are incorporated in the DMFSGD algorithm by default.

Second, $\lambda = 1$ seems to be a good choice under most configurations and is thus adopted by default. Third, the impact of $r$ depends on the properties of the dataset. In Harvard226 where available measurements are dense, the prediction accuracy improves monotonically with $r$, whereas in the other two datasets where available measurements are sparse due to the setting of a small $k$, better performance is achieved with $r \leqslant 10$ and a large $\lambda$ is needed to overcome the overfitting caused by larger $r$'s, which confirms our analysis in the previous section. Thus, by trading off between the performance on all three datasets, $r = 10$ is adopted by default.

*3) $\eta$:* As mentioned earlier, SGD is sensitive to the learning rate $\eta$ where a too large $\eta$ leads to the overflow of the solutions and a too small $\eta$ slows down the convergence. Although this sensitivity is reduced by minibatch SGD, it is still difficult to find an appropriate constant that works for all datasets and in all situations. We experimented with different constant $\eta$'s and with the line search to adapt $\eta$ dynamically. Results are shown in Figure 7. It can be seen that the line search strategy performs best in terms of both accuracy and convergence speed. Note that the convergence speed is illustrated by the stress and MAE improvements with respect to the average measurement number per node, i.e. the total number of measurements used by all nodes divided by the number of nodes[3]. It can be seen that the DMFSGD algorithm converges fast after each node probe, on average, $10 \times k$ measurements from its $k$ neighbors. Although no $k$ is set for Harvard226, we treat it as $k = 226$.

*4) Discussions:* By incorporating the line search strategy, the $L_1$ loss function and the nonnegativity constraint, our DMFSGD algorithm is left with two tunable parameters: the rank $r$ and the regularization coefficient $\lambda$. The default configuration of $\lambda = 1$ and $r = 10$ is not guaranteed to be optimal in different situations and on different datasets. However, fine tuning of parameters is difficult, if not impossible, for network applications due to the measurement dynamics and the decentralized processing where local measurements are processed locally at each node with no central nodes gathering information of the entire network. Empirically, the default parameter setting leads to good, though not the best, prediction accuracy to a large variety of data.

The setting of $k = 32$ has been commonly adopted in many systems such as Vivaldi. However, most systems contain network nodes of a few thousands or less. For large systems of more nodes, $k$ has to be scaled with the number of nodes $n$. According to the theory of matrix completion [19]–[21], one can recover an unknown $n \times n$ matrix of low rank $r$ from just about $O(nr\log n)$ noisy entries with an error which is proportional to the noise level. Thus, $k \propto r\log n$ to guarantee a decent prediction accuracy.

### D. Comparisons with Vivaldi

Among numerous approaches on network distance prediction, we consider Vivaldi [9] as the state of the art because of its accuracy and its practicability. To the best of our knowledge, Vivaldi is the only system that has been actually adopted in a real application, namely Azureus [2]. Other approaches such as GNP [8] and IDES [9] are less convenient due to the usage of landmarks, which makes their application impossible in the context of passive probing of distance measurements (thus impossible to be evaluated on the Harvard226 dataset). Due to the insights in Section III-D, we consider these landmark-based systems as a special variation of a generic decentralized model.

In this paper, we only compare our DMFSGD algorithm with Vivaldi. To address the measurement dynamics and the skewed neighbor updates, we adopted the Vivaldi implementation in [13] [4] when dealing with the Harvard226 dataset. The conventional Vivaldi in [9] was adopted to deal with the other two datasets. We refer to the former as Harvard Vivaldi to make the distinction. In addition, despite the impracticality, we also demonstrate the flexibility of the DMFSGD algorithm in dealing with the landmark-based architecture, referred to as DMFSGD Landmark, by forcing each node to only select the landmarks as neighbors. Note that we only ran DMFSGD Landmark on P2PSim1740 and Meridian2500 because the dynamic measurements in Harvard226 were obtained passively and thus we cannot select landmarks and force each node to only communicate with them. To make the comparison fair, 32 landmarks were randomly selected.

[3]For P2PSim1740 and Meridian2500, at any time, the number of measurements used by each node is statistically the same for all nodes due to the random selections of the source and the target nodes in the updates. For Harvard226, this number is significantly different for different nodes because the paths were passively probed with uneven frequencies.

[4]The source code was downloaded from http://www.eecs.harvard.edu/~syrah/nc/.

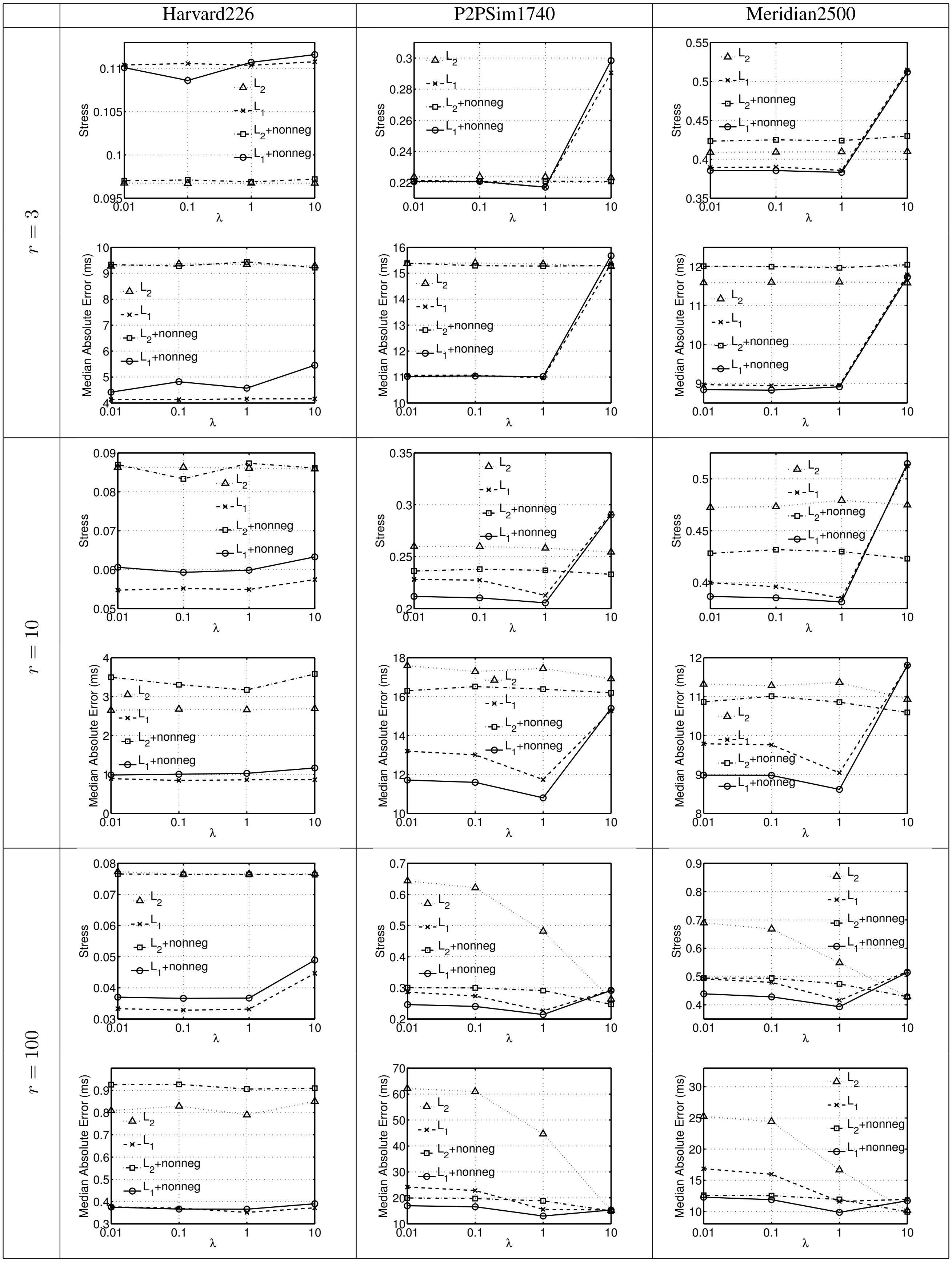

Fig. 6. Impact of parameters. $\eta$ is adapted by the line search.

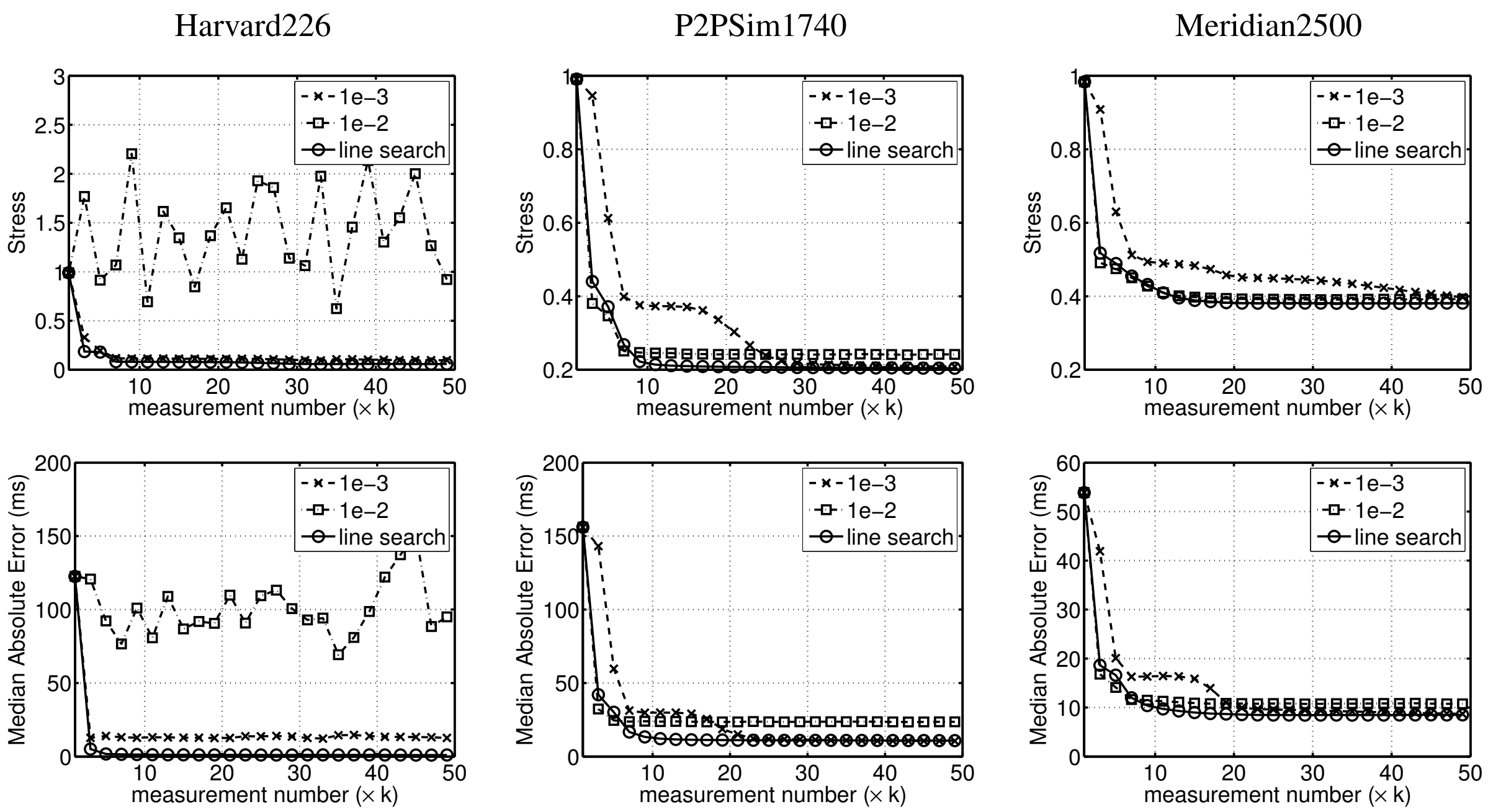

Fig. 7. Impact of $\eta$ under $\lambda = 1$ and $r = 10$ with the $L_1$ loss function and the nonnegativity constraint. $k$ is treated as 226 for Harvard226 and $k = 32$ for P2PSim1740 and Meridian2500.

Figure 8 shows the comparisons between DMFSGD, Vivaldi/Harvard Vivaldi and DMFSGD Landmark. It can be seen that on different criteria, our DMFSGD algorithm either outperforms or is competitive with Vivaldi. On Harvard226, DMFSGD is significantly better on all criteria, especially on the MAE where DMFSGD achieved the $1ms$ MAE, in contrast to the $5ms$ by Harvard Vivaldi, meaning that half of the estimated distances have an error of less than $1ms$. On P2PSim1740, DMFSGD is better on the MAE and the cumulative distributions of REE, whereas on Meridian2500, DMFSGD achieved similar performance as Vivaldi on all criteria. Note that DMFSGD and DMFSGD Landmark performed similarly on P2PSim1740 and Meridian2500.

As Harvard226 contains real dynamic data collected from a real application, Azureus, the superiority on it shows clearly the usability of our DMFSGD algorithm.

## VII. Conclusions and Future Works

This paper presents a novel approach to network distance prediction by low-rank matrix factorization. The success of the approach roots both in the exploitation of the dependencies across distance measurements between network nodes and in the stochastic optimization which enables a fully decentralized architecture. A so-called Decentralized Matrix Factorization by Stochastic Gradient Descent (DMFSGD) algorithm is proposed to solve the distance prediction problem. The algorithm is simple, with no infrastructure, scalable, able to deal with dynamic measurements in large-scale networks, and accurate, generally superior to Vivaldi.

Extensive experiments on various RTT datasets, particularly on one with real data from Azureus, demonstrate the potential of the algorithm being utilized by Internet applications, which we would like to study in the future. Our approach is flexible and can easily be extended to other network metrics such as available bandwidth [43]. An interesting topic is to study which metrics are suitable for our matrix completion framework.

## Acknowledgments

This work was partially supported by the EU under project FP7-Fire ECODE, by the European Network of Excellence PASCAL2 and by the Belgian network DYSCO (Dynamical Systems, Control, and Optimization), funded by the Interuniversity Attraction Poles Programme, initiated by the Belgian State, Science Policy Office. The scientific responsibility rests with its authors.

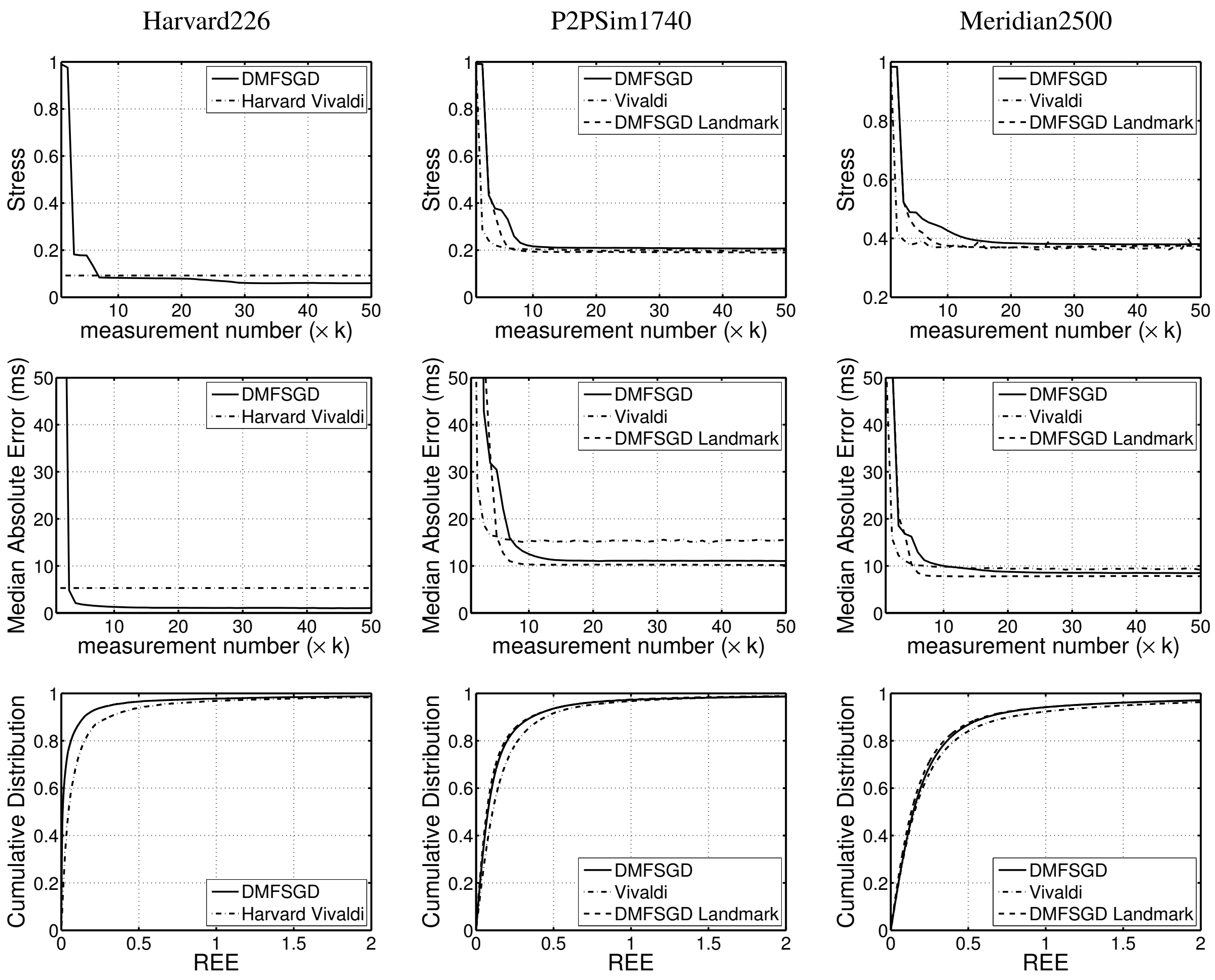

Fig. 8. Comparison of DMFSGD and Vivaldi. The default configuration of $\lambda = 1$ and $r = 10$ with $\eta$ adapted by the line search, the $L_1$ loss function and the nonnegativity constraint is used in DMFSGD and DMFSGD Landmark. The 10 dimensional Euclidean space with the Height model is used in Vivaldi and Harvard Vivaldi. Note that as the implementation of Harvard Vivaldi only outputs the results in the end of the simulation, the final stress and the final MAE are plotted as a constant.

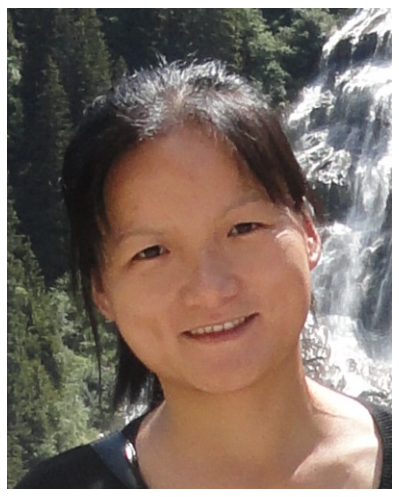

**Yongjun Liao** is a PhD student at Department of Electrical Engineering and Computer Science, University of Liège, Belgium. She received her B.S. in 1999 and M.S. in 2002, both from Department of Computer Science, Guangxi University, China. Before joining the networking group RUN in University of Liège in 2007, she had worked as a software engineer in a small computer company in Beijing, China. Her main research interests are the applications of machine learning techniques to computer networking problems, specifically the prediction of end-to-end network performance in large-scale networks.

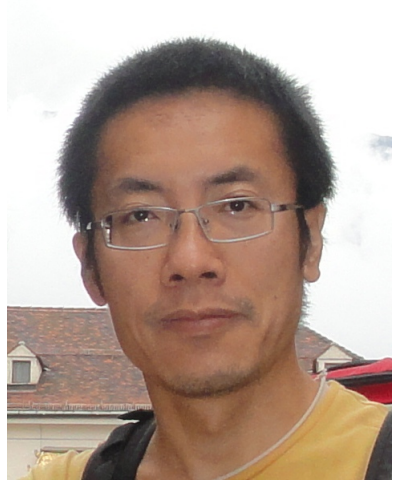

**Wei Du** is a senior Postdoctoral researcher at Intelligent and Interactive Systems (IIS), University of Innsbruck, Austria. He received his B.S. in 1997 from Tianjin University, China, and PhD in 2002 from Institute of Computing Technology, Chinese Academy of Sciences, China. Since graduation, he has been working as postdoctoral researcher at INRIA, France, Hamburg University, Germany, University of Liège, Belgium, and University of Innsbruck, Austria. His main research interests are computer vision and machine learning.

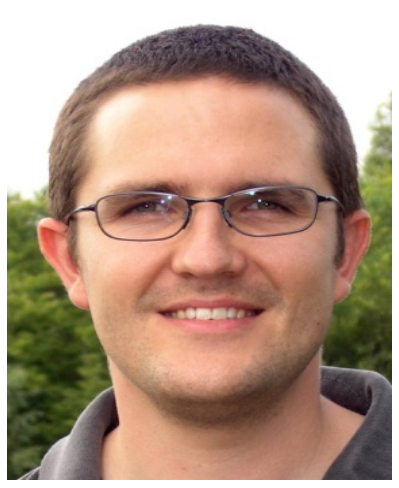

**Pierre Geurts** is an assistant professor in the EECS department of the University of Liège, Belgium. He graduated as an electrical (computer science) engineer in 1998 and received the PhD degree in applied sciences in 2002. From 2006 to 2011, he was research associate of the FNRS (Belgium). His research interests concern the design of, computationally and statistically efficient, supervised and semi-supervised learning algorithms in order to exploit structured input and output spaces (sequences, images, time-series, graphs), with applications in bioinformatics, computer vision, and computer networks.

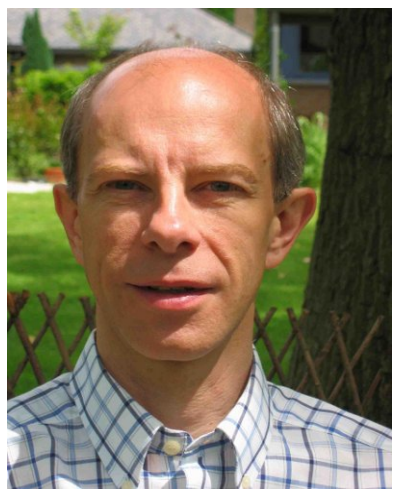

**Guy Leduc** is a full professor in the EECS department of the University of Liège, Belgium, and is since 1997 the head of the Research Unit in Networking (RUN). He graduated as an electrical (electronics) engineer in 1983 and got his PhD in computer science in 1991.

His research field is computer networks, and his main research interests are Network Coordinate Systems, overlays, traffic engineering, resilience, multimedia, congestion control, and autonomic/active/programmable networks. His research unit is or has been involved in European projects such as ECODE on cognitive networking, ResumeNet on Resilient Networking, ANA on autonomic networking, TOTEM on an open-source toolbox for traffic engineering, and the E-NEXT European network of excellence.

Since 2007 he has been the chairman of the IFIP Technical Committee (TC6) on Communications Systems. He is an area editor of the Elsevier Computer Communications journal, and a steering committee member of the IFIP Networking Conference.